\documentclass[journal]{IEEEtran}
\IEEEoverridecommandlockouts
\usepackage{cite}
\usepackage[colorlinks=true,linkcolor=black,anchorcolor=black,citecolor=black,filecolor=black,menucolor=black,runcolor=black,urlcolor=blue]{hyperref}

\usepackage{amsmath,amssymb,amsfonts,mathrsfs}
\usepackage[defaultcolor=red]{changes}
\usepackage{algorithmic}
\usepackage{graphicx}
\usepackage{textcomp}
\usepackage{xcolor}
\usepackage{tabularx,booktabs}
\newcolumntype{C}{>{\centering\arraybackslash}X} 
\def\BibTeX{{\rm B\kern-.05em{\sc i\kern-.025em b}\kern-.08em
    T\kern-.1667em\lower.7ex\hbox{E}\kern-.125emX}}

\ifCLASSOPTIONcompsoc
\usepackage[caption=false,font=normalsize,labelfont=sf,textfont=sf]{subfig}
\else
\usepackage[caption=false,font=footnotesize]{subfig}
\fi

\newcommand{\argmin}[1]{\underset{#1}{\operatorname{argmin}}\;}

\newcommand{\bs}[1]{\boldsymbol{#1}}

\newcommand{\mob}[2]{\mathcal{M}_{#1}\left({#2}\right)}

\begin{document}
\bstctlcite{Settings}
\title{Automatic Model Extraction of the Match Standard in Symmetric--Reciprocal--Match Calibration}

\author{%
	\IEEEauthorblockN{%
        Ziad~Hatab, Michael~E.~Gadringer, Arash~Arsanjani, and~Wolfgang~B\"osch%
    }%
    \thanks{%
        Ziad~Hatab was with the Institute of Microwave and Photonic Engineering, 
        Graz University of Technology, 8010 Graz, Austria, during the development 
        of this work (e-mail: z.hatab@alumni.tugraz.at).%
        \par Michael~E.~Gadringer, Arash~Arsanjani, and~Wolfgang~B\"osch are with 
        the Institute of Microwave and Photonic Engineering, Graz University of 
        Technology, 8010 Graz, Austria, and with the Christian Doppler Laboratory 
        for Technology-Guided Electronic Component Design and Characterization 
        (TONI), 8010 Graz, Austria (e-mail: \{michael.gadringer, arash.arsanjani, 
        wbosch\}@tugraz.at).%
        \par This work was supported by the Christian Doppler Research Association, the Austrian Federal Ministry for Digital and Economic Affairs, and the National Foundation for Research, Technology and Development.%
        \par Software implementation and measurement data are available at:\\ 
        \url{https://github.com/ziadhatab/srm-calibration}%
    }%
}%
\markboth{This work has been accepted for publication in the IEEE Transactions on Instrumentation and Measurement}{}
\maketitle

\begin{abstract}
This paper addresses the modeling of parasitics of the match standard in the symmetric-reciprocal-match (SRM) calibration method of vector network analyzers (VNAs). In the original SRM procedure, the match standard is assumed to be defined. Here, we demonstrate that the match can be modeled with an arbitrary frequency-dependent model using a non-linear global optimization procedure. To highlight the validity of the suggested approach, numerical tests were conducted, demonstrating the ability to recover the match standard parasitic model down to software numerical precision. Additionally, we performed microstrip line measurements on PCB to compare the SRM calibration using the proposed automatic model extraction against an SRM calibration where the match standard is explicitly defined via multiline thru-reflect-line (TRL) data. The results show that the automatic model extraction can achieve accuracy comparable to that of a data-based match definition, validating the approach as an effective substitute when explicit characterization of the match standard is unavailable.
\end{abstract}


\section{Introduction}
\label{sec:1}

\IEEEPARstart{T}{he} calibration of vector network analyzers (VNAs) is an essential step to compensate for the systematic imperfections of the measurement equipment and the measurement setup. The most widely known calibration procedure for VNAs is the short-open-load-thru (SOLT) method\,\cite{Kruppa1971}, which requires all four standards to be fully characterized. The short-open-load-reciprocal (SOLR) method\,\cite{Ferrero1992} extends SOLT by replacing the thru with any transmissive reciprocal device. However, both methods are fundamentally limited by the accuracy of the standard definitions, which can be challenging to achieve across broad frequency ranges and different measurement environments. To address this, self-calibration procedures emerged that rely on partially defined standards\,\cite{Rumiantsev2008}.

Among the most accurate self-calibration methods is multiline TRL\,\cite{Engen1979,Marks1991,Hatab2022,Hatab2023}, which depends on distributed circuit models rather than lumped-element approximations, making it traceable and suitable as a primary calibration\,\cite{Wong2008}. However, multiline TRL becomes impractical at lower frequencies due to extremely long line standards and requires variable distances between measurement planes.

For on-wafer applications, the line-reflect-match (LRM), thru-match-reflect-reflect (TMRR), and line-reflect-reflect-match (LRRM) methods\,\cite{Eul1988,Zhao2017,Rumiantsev2018,Hayden2006} are commonly used. The LRRM method\,\cite{Hayden2006} introduced series inductance modeling for the match standard parasitics, later extended to include a transmission line section and a shunt capacitance\,\cite{Liu2017}. However, these simplified models cannot accurately represent various resistor implementations, such as via-based shunt resistors or surface-mounted resistors with parallel capacitance, which are scenarios common in device characterization at the printed circuit board (PCB) level. Furthermore, LRM/TMRR/LRRM require defined line standards for reference-plane positioning, making them challenging when measurement contacts, whether probes or connector launchers, are positioned orthogonally or at angles, as bent lines are more challenging to define\,\cite{Basu1997}.

Alternative approaches to calibration include the series-resistor method\,\cite{Williams1997,Liu2016,Drisko2020}, demanding minimal parasitics and high manufacturing precision for on-wafer applications, and Bayesian optimization methods\,\cite{Hoffmann2009,Zhou2023} that physically model all standards. The latter approach is computationally intensive and requires modeling every standard, whereas our method limits modeling to the match standard while exploiting symmetry and reciprocity for the remaining standards.

The symmetric-reciprocal-match (SRM) method\,\cite{Hatab2024} combines the advantages of LRRM/LRM/TMRR (undefined symmetric standards) with SOLR (unknown transmissive reciprocal device), requiring only a defined match standard. This is particularly advantageous for PCB environments where standard definition inaccuracies are common\,\cite{Hatab2023d,Takahashi2023}. In the original publication on SRM calibration\,\cite{Hatab2024}, the procedure was demonstrated using coaxial measurements. However, it can also be applied to waveguide scenarios, as defining the measurement standards through symmetry and reciprocity can be easily accomplished with waveguide shims and shorts.

This paper extends SRM by incorporating automatic parasitic extraction of the match standard, requiring only its DC resistance while modeling the frequency-dependent parasitic response. Unlike LRRM's simplified models\,\cite{Hayden2006,Liu2017}, our approach employs global optimization to accommodate complex structures, which can include effects like via hole parasitics in shunt resistor implementations. While this work focuses primarily on quasi-TEM microstrip structures on PCBs, where diverse parasitics commonly arise from circuit topology, the method can be extended to non-TEM scenarios, such as waveguides. In such a non-TEM case, instead of relying on DC measurement to anchor the definition of the match, waveguide parameter models are optimized\,\cite{Lomakin2018}.

To validate the proposed approach, we conducted numerical tests to verify the optimization procedure's ability to recover the underlying model parameters, demonstrating that the match standard parasitic model can be extracted down to software numerical precision. Additionally, we performed PCB microstrip line measurements to compare the SRM calibration obtained using the proposed automatic model extraction with an SRM calibration in which the match standard is explicitly defined via multiline TRL characterization. The results demonstrate that automatic model extraction achieves accuracy comparable to using a data-based match definition, validating the approach as an effective alternative when explicit characterization of the match standard is unavailable.

The article is organized as follows. Section~\ref{sec:2} reviews the general procedure for the SRM method. Section~\ref{sec:3} discusses the parasitic modeling of the match and the optimization procedure. Section~\ref{sec:4} presents numerical analysis using synthetic data and experimental measurements on PCBs, followed by conclusions in Section~\ref{sec:5}.

\section{SRM Calibration Review}
\label{sec:2}

\begin{figure}[th!]
    \centering
    \includegraphics[width=0.95\linewidth]{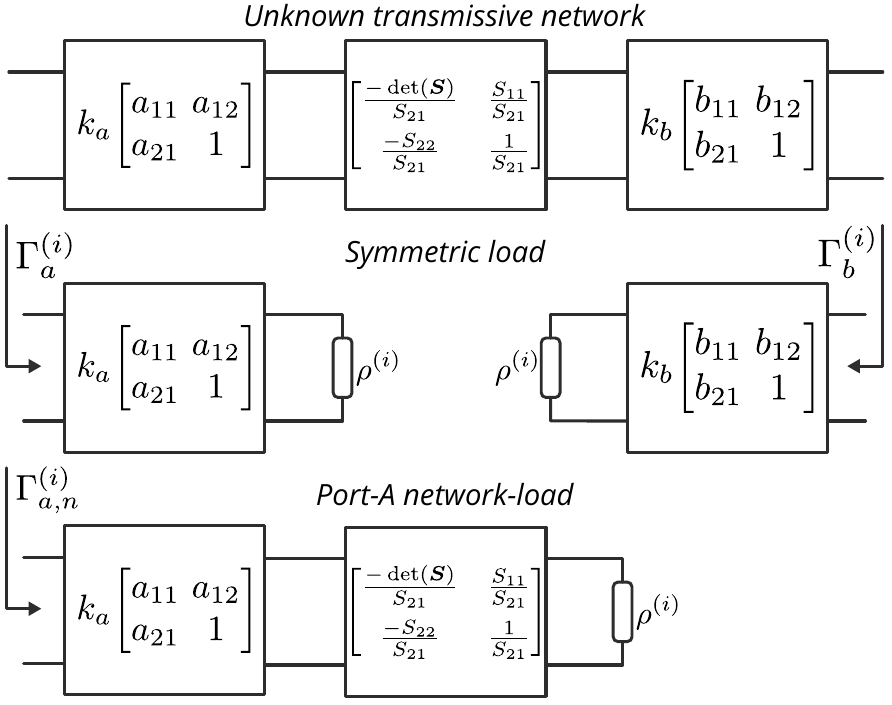}
    \caption{Two-port VNA error box model illustrating the standards used in SRM calibration. All matrices are provided as T-parameters. The index $i$ indicates the measured standard, where $i=1,2,\ldots, M$, with $M \geq 3$. Note that the last standard could also be measured at port-B.}
    \label{fig:2.1}
\end{figure}

The SRM calibration method relies on measuring symmetric one-port standards without predefined characteristics, combined with a transmissive device measurement. This combination enables the formulation of an eigenvalue problem that partially solves for the VNA error boxes. The complete solution is then obtained using the reciprocity property of the transmissive device and a pair of defined match standards. The method requires a minimum of three symmetric one-port standards (typically open, short, and an arbitrary load), measured both individually and in combination with the transmissive device from either port, as illustrated in Fig.~\ref{fig:2.1}.

The measurement of the transmissive device can be written in T-parameters as follows:
\begin{equation}
    \bs{M}_\mathrm{net} = \underbrace{k_ak_b}_{k}\underbrace{\left[\begin{matrix}a_{11} & a_{12}\\
            a_{21} & 1\end{matrix}\right]}_{\bs{A}}\bs{T}_\mathrm{net} \underbrace{\left[\begin{matrix}b_{11} & b_{12}\\
            b_{21} & 1\end{matrix}\right]}_{\bs{B}}, 
    \label{eq:2.1}
\end{equation}
where $\bs{M}_\mathrm{net}$ and $\bs{T}_\mathrm{net}$ represent the measured and actual T-parameters of the network, respectively. The matrices $\bs{A}$ and $\bs{B}$ represent the error boxes of the left and right ports, respectively, containing the first six error terms. The parameter $k$ identifies the seventh error term, which describes the transmission error between the ports.

The measurements of the one-port standards can be represented using M\"{o}bius transformation notation, as described below \cite{Speciale1981,Hatab2024}:\begin{subequations}
    \begin{align}
        \mob{\bs{A}}{\rho^{(i)}} &= \Gamma_a^{(i)} = \frac{a_{11}\rho^{(i)} + a_{12}}{a_{21}\rho^{(i)} + 1},\label{eq:2.2a}\\
        \mob{\bs{P}\bs{B}^{-1}\bs{P}}{\rho^{(i)}} &= \Gamma_b^{(i)} = \frac{b_{11}\rho^{(i)} - b_{21}}{1 - b_{12}\rho^{(i)}}\label{eq:2.2b},
    \end{align}
    \label{eq:2.2}
\end{subequations}
where $\Gamma_a^{(i)}$ and $\Gamma_b^{(i)}$ are the $i$th measured reflection coefficients from the left and right ports, respectively. The actual response of the standard is denoted by $\rho^{(i)}$. The matrix $\bs{P}$ is a $2\times 2$ permutation matrix,
\begin{equation}
    \bs{P} = \bs{P}^T = \bs{P}^{-1} = \begin{bmatrix}
        0 & 1\\
        1 & 0
    \end{bmatrix}.
    \label{eq:2.3}
\end{equation}

A M\"{o}bius transformation of the form $\mob{\bs{Q}}{z}$ takes the elements of a $2\times 2$ matrix $\bs{Q}$ and maps the input variable $z$ through a linear fractional transformation. This is written in general as follows: 
\begin{equation}
    \mob{\bs{Q}}{z} = \frac{q_{11}z + q_{12}}{q_{21}z + q_{22}}; \qquad \bs{Q} = \begin{bmatrix}
        q_{11} & q_{12}\\ q_{21} & q_{22}
    \end{bmatrix}
    \label{eq:2.4}
\end{equation}

The reason for expressing the one-port measurements as M\"{o}bius transformations is that the operations can be composed, which translates to a simple matrix product for the coefficients \cite{Needham2023}. This allows us to easily express the inverse of the operation by taking the inverse of the matrix. For example, for the port-B measurement from \eqref{eq:2.2b}, the input coefficient can be reverse calculated as follows:
\begin{equation}
    \mob{\bs{P}\bs{B}\bs{P}}{\Gamma_b^{(i)}}  = \frac{\Gamma_b^{(i)}+b_{21}}{b_{12}\Gamma_b^{(i)}+b_{11}} = \rho^{(i)}.
    \label{eq:2.5}
\end{equation}

Further details on M\"{o}bius transformation and its properties can be found in\,\cite{Needham2023}. What is important here is that the measurement from port-A can be combined with the measurement from port-B, as they measure the same standard $\rho^{(i)}$ due to the imposed symmetry property, as follows:
\begin{equation}
    \Gamma_a^{(i)} = \mob{\bs{A}}{\mob{\bs{P}\bs{B}\bs{P}}{\Gamma_b^{(i)}}} = \mob{\underbrace{\bs{A}\bs{P}\bs{B}\bs{P}}_{\bs{H}}}{\Gamma_b^{(i)}}
    \label{eq:2.6}
\end{equation}

Therefore, with at least three symmetric one-port measurements, it is possible to retrieve the coefficients of the matrix product $\bs{H} = \bs{A}\bs{P}\bs{B}\bs{P}$ through a least-squares or nullspace formulation \cite{Hatab2024}. 

Similarly to \eqref{eq:2.6}, another M\"{o}bius transformation composition can be written using the one-port measurement cascaded with the transmission network at port-A with the measurement at port-B as follows:
\begin{equation}
    \Gamma_{a,n}^{(i)} = \mob{\underbrace{\bs{A}\bs{T}_\mathrm{net}\bs{P}\bs{B}\bs{P}}_{\bs{F}}}{\Gamma_b^{(i)}},
    \label{eq:2.7}
\end{equation}
where $\bs{F}$ is also solved using a least-squares or nullspace solution.

Therefore, with $\bs{H}$ and $\bs{F}$ and the measurement of the unknown transmissive device $\bs{M}_\mathrm{net}$ from \eqref{eq:2.1}, it can be shown that a ``virtual'' thru standard can be created that can be used to formulate an eigenvalue problem for both port-A and port-B coefficients. Reference \cite{Hatab2024} discusses this in greater detail and provides the equations for the case of measuring the cascaded one-port standards at port-B. 

In general, the structure of the obtained eigenvalue problem at each port is given by the following expressions:
\begin{subequations}
    \begin{align}
        \frac{k}{\nu} \bs{A}\bs{P}\bs{A}^{-1} &= \bs{W}_a\bs{\Lambda}\bs{W}_a^{-1},\quad \forall k,\nu\neq 0 \label{eq:2.8a}\\
        \frac{k}{\nu} \bs{B}^T\bs{P}\bs{B}^{-T} &= \bs{W}_b\bs{\Lambda}\bs{W}_b^{-1},\quad \forall k,\nu\neq 0\label{eq:2.8b}
    \end{align}
    \label{eq:2.8}
\end{subequations}
where $\bs{W}_a$ and $\bs{W}_b$ are the eigenvectors to be solved for, and $\bs{\Lambda}$ contain the eigenvalues.

The last step is to incorporate the defined match standard to formulate a nullspace problem, as shown below:
\begin{equation}
    \left[\begin{matrix}
        -1 & -1 & w_{11}^{(a)} & w_{11}^{(a)} \\
        1 & -1 & -w_{12}^{(a)} & w_{12}^{(a)} \\
        -\rho_a^{(m)} & -1 & \Gamma_a^{(m)}\rho_a^{(m)} & \Gamma_a^{(m)}
    \end{matrix}\right]
    \left[\begin{matrix}
        a_{11}\\
        a_{12}\\
        a_{21}\\
        1
    \end{matrix}\right] = \bs{0}
    \label{eq:2.9}
\end{equation}

The system of equations for port-B can be obtained similarly, resulting in the following system of equations:
\begin{equation}
    \left[\begin{matrix}
        -1 & -1 & w_{11}^{(b)} & w_{11}^{(b)} \\
        1 & -1 & -w_{12}^{(b)} & w_{12}^{(b)} \\
        -\rho_b^{(m)} & 1 & -\Gamma_b^{(m)}\rho_b^{(m)} & \Gamma_b^{(m)}
    \end{matrix}\right]
    \left[\begin{matrix}
        b_{11}\\
        b_{21}\\
        b_{12}\\
        1
    \end{matrix}\right] = \bs{0}
    \label{eq:2.10}
\end{equation}
where $\rho_a^{(m)}$ and $\rho_b^{(m)}$ are the defined match standards at both ports (they do not need to be equal), and $\Gamma_a^{(m)}$ and $\Gamma_b^{(m)}$ are their corresponding measurements, respectively. The coefficients $w_{ij}^{(a)}$ and $w_{ij}^{(b)}$ are the coefficients from the eigenvectors for both ports. Lastly, the final error term $k$ is solved using the unknown-thru procedure as in the SOLR procedure \cite{Ferrero1992}.

In the entire process of solving for the error terms, only the match standard needs to be explicitly defined, while all other standards are characterized through symmetry and reciprocity. The discussion in Section~\ref{sec:3} addresses how to solve \eqref{eq:2.9} and \eqref{eq:2.10} without explicitly defining the match standard, but instead through an equivalent circuit modeling to capture its parasitic response.


\section{Automatic Parasitic Model Extraction}
\label{sec:3}

In the derivation outlined in Section~\ref{sec:2}, \eqref{eq:2.9} and \eqref{eq:2.10} were obtained. These equations can only be solved if a fully defined match standard is used. In many cases, defining the match standard accurately at higher frequencies can be challenging, as parasitic elements become more significant as the wavelength approaches the device's physical dimensions. This is further demonstrated with measurements in Section~\ref{sec:4}.

\begin{figure}[th!]
    \centering
    \includegraphics[width=1\linewidth]{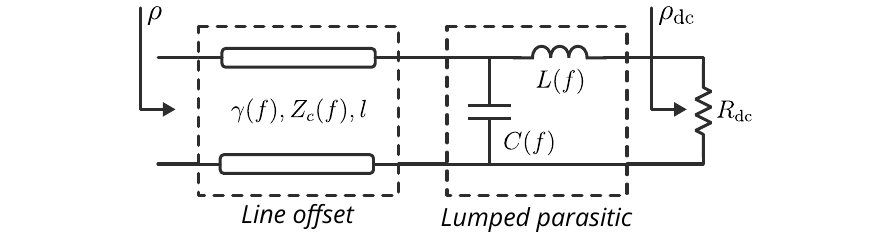}
    \caption{Equivalent circuit model of a non-ideal match standard. The values of $l$ and $\rho_\mathrm{dc}$ are assumed to be known. The parameters $L(f)$, $C(f)$, $\gamma(f)$, and $Z_c(f)$ are modeled as frequency-dependent.}
    \label{fig:3.1}
\end{figure}

In general, the parasitic behavior of the match standard is more complex than just a series inductance or shunt capacitance. These non-idealities can come as a mixture of series and shunt behavior and may even include a transmission line segment. An example of such a model is depicted in Fig.~\ref{fig:3.1}. In this model, the DC resistance of the match standard is known, as well as the length of the offset. The remaining parameters are modeled as frequency-dependent with unknown constants. The lumped elements $L(f)$ and $C(f)$ can be modeled using polynomials, whereas the transmission line parameters, i.e., propagation constant $\gamma(f)$ and characteristic impedance $Z_c(f)$, can be modeled using corresponding analytical or semi-analytical equations of the physical cross-section of the transmission line.

It should be noted that the DC resistance needs to be defined, as otherwise, it would not be possible to define the reference impedance. With knowledge of the DC resistance, the definition of the match standard at DC can be anchored. Also, the length of the transmission line needs to be defined for the same reason, as otherwise, a scaling ambiguity in the definition of $\gamma(f)$ would arise. 

While the equivalent circuit model presented in Fig.~\ref{fig:3.1} is an example of a match standard, it could also be used to describe non-ideal open or short standards, where $\rho_\mathrm{dc}=1$ is for an open standard and $\rho_\mathrm{dc}=-1$ is for a short standard. However, in general, the model's complexity depends on the physical design and manufacturing technology of the match standard. For example, in the measurement example discussed in Section~\ref{sec:4b}, a flip-chip resistor is used, which needed to be described with a more complex lumped model to account for the mounting pads and the resistor's internal parasitics.

The question is: if a one-port device can be described by an equivalent circuit model, how can the unknown parameters be solved without the match standard's measurement? This can be addressed through non-linear optimization of the nullspace problem through singular value decomposition (SVD). Recalling the system of equations earlier in either \eqref{eq:2.9} or \eqref{eq:2.10}, this system can be expanded by including other one-port standards besides the match standard, as shown in \eqref{eq:3.2} below for port-A:
\begin{equation}
    \underbrace{\left[\begin{matrix}
            -1 & -1 & w_{11}^{(a)} & w_{11}^{(a)} \\
            1 & -1 & -w_{12}^{(a)} & w_{12}^{(a)} \\
            -\rho_a^{(i)}(\bs{\theta}_i) & -1 & \Gamma_a^{(i)}\rho_a^{(i)}(\bs{\theta}_i) & \Gamma_a^{(i)}\\
            \vdots & \vdots & \vdots & \vdots
        \end{matrix}\right]}_{\bs{G}_a(\bs{\theta})}
    \left[\begin{matrix}
        a_{11}\\
        a_{12}\\
        a_{21}\\
        1
    \end{matrix}\right] = \bs{0}
    \label{eq:3.2}
\end{equation}
where $\Gamma_a^{(i)}$ and $\rho_a^{(i)}(\bs{\theta}_i)$ represent the raw measurement and reflection coefficient, respectively, of the $i$th load standard. The vector $\bs{\theta}_i$ contains all unknown frequency-independent parameters that need to be determined, with $\bs{\theta} = [\bs{\theta}_1^T, \bs{\theta}_2^T, \ldots]^T$. It is important to note that the match standard must be part of the $i$th standards, along with at least one other one-port standard, such as a short, open, or arbitrary non-match impedance, or all of them simultaneously. Therefore, $i>1$ is required to have an overdetermined system of equations.

Since $\bs{G}_a(\bs{\theta})$ must have a rank of 3 to be solvable, and its nullspace corresponds to the solution vector, the optimal choice of $\bs{\theta}$ should result in a minimal fourth singular value (ideally zero), which corresponds to the nullspace \cite{Strang1993}. Therefore, the fourth singular value corresponding to the nullspace of $\bs{G}_a(\bs{\theta})$ can be calculated at each frequency point and used to determine the parameter vector $\bs{\theta}$ that minimizes this value across all frequency points. This optimization procedure over the parameter vector $\bs{\theta}$ can be expressed as minimizing the average fourth singular value across all frequency points, described as follows:
\begin{equation}
    \bs{\theta}_\mathrm{opt} = \argmin{\bs{\theta}} \frac{1}{N}\sum_{i=1}^{N}\sigma_4(\bs{\theta}, f_i)
    \label{eq:3.3}
\end{equation}
where $f_i$ refers to the $i$th frequency point, and $N$ is the total number of frequencies. Once $\bs{\theta}_\mathrm{opt}$ has been found, the parameters associated with the match standard are used to define its reflection coefficient accurately. 

The optimization procedure in \eqref{eq:3.3} can only be solved when dealing with an overdetermined system of equations, which requires having at least one additional measurement besides the match standard. This can be either a short or open standard that was assumed unknown during calibration. These standards are modeled similarly to the match standard. For the optimization itself, since singular values do not necessarily form a convex problem, a bounded global optimization method, e.g., differential evolution (DE) \cite{Storn1997} needs to be used.

After $\bs{\theta}_\mathrm{opt}$ is found, the coefficients associated with the match standard are substituted back into \eqref{eq:2.9} and \eqref{eq:2.10}. If using symmetric match pairs, optimization can be performed on only one side. However, for asymmetric match standards, each port must be solved independently. Furthermore, while additional unknown standards can be included in the SVD procedure (e.g., both short and open standards), this significantly increases the number of parameters to be solved, as the unknown parameters of each standard are not coupled. This makes it challenging even for global optimization methods to find all parameters simultaneously. Therefore, it is best practice to minimize the complexity of the optimization whenever possible, either through the number of solved parameters or by coupling the parameters across both ports. Additionally, the speed of the optimization is also a function of the evaluation of the models. For example, incorporating a transmission line segment, as in the numerical example in Section~\ref{sec:4a}, can be computationally expensive when using special integral functions as the Bessel functions \cite{Heinrich1993}.

\section{Experiments}
\label{sec:4}

This section discusses two experiments. The first experiment involves numerical analysis using synthetic data to demonstrate the accuracy of fitting the match standard using a nonlinear optimization procedure compared to using an ideal definition for the match standard. The second experiment involves PCB measurements to validate the automatic parasitic model extraction procedure against multiline TRL calibration. 

\subsection{Numerical Analysis}
\label{sec:4a}
\begin{figure}[th!]
    \centering
    \includegraphics[width=0.95\linewidth]{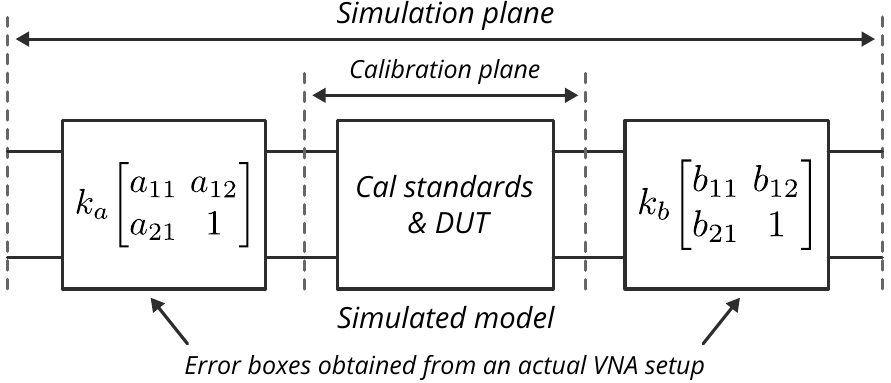}
    \caption{Block diagram illustration of the numerical simulation concept to generate realistic synthetic data.}
    \label{fig:4.1}
\end{figure}

The numerical analysis involves creating synthetic data of co-planar waveguide (CPW) standards using the model developed in \cite{Heinrich1993,Schnieder2003,Phung2021a}. To accurately simulate an on-wafer setup, error boxes extracted from actual on-wafer measurements are utilized using multiline TRL calibration on a CPW impedance substrate standard (ISS). For detailed information about the measurement setup, calibration substrate, and measurements, readers are referred to \cite{Hatab2023}. However, relevant information about the CPW parameters is presented below. The goal is to generate SRM standards based on the CPW model and embed them within the error boxes of an actual VNA setup. Fig.~\ref{fig:4.1} illustrates the experimental setup.

The simulated CPW cross-section structure has the following dimensions, which were taken from \cite{Hatab2023}: signal width of $49.1\,\mu\mathrm{m}$, ground width of $273.3\,\mu\mathrm{m}$, conductor spacing of $25.5\,\mu\mathrm{m}$, and conductor thickness of $4.9\,\mu\mathrm{m}$. The substrate is Alumina with a real-part relative permittivity of $\epsilon_r^\prime=9.9$ and loss tangent of $\tan\delta=0.0002$. The conductor is gold with relative conductivity to copper of $\sigma_r=0.7$, where copper conductivity is $\sigma=58\,\mathrm{MS/m}$. It should be noted that the Alumina substrate considered in \cite{Hatab2023} was assumed to have a loss tangent of zero, as the manufacturer does not provide a value and states it is lossless. We assumed a small $\tan\delta$ of 0.0002 for the simulation, as almost all materials exhibit some level of loss at room temperature.

\begin{figure}[th!]
    \centering
    \subfloat[]{\includegraphics[width=.49\linewidth]{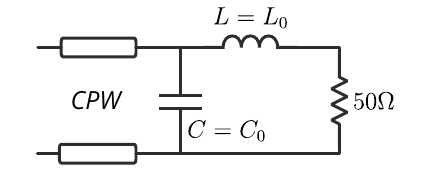}}\label{fig:4.2a}
    \subfloat[]{\includegraphics[width=.49\linewidth]{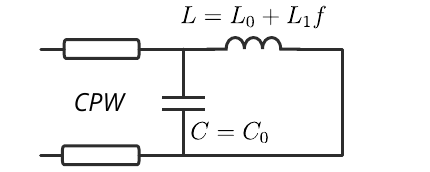}}\label{fig:4.2b}\\[-1pt]
    \subfloat[]{\includegraphics[width=.49\linewidth]{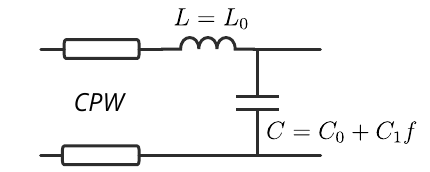}}\label{fig:4.2c}
    \caption{Models used to simulate non-ideal load standards (a) $50\,\Omega$ match standard with $L_0=25\,\mathrm{pH}, C_0=1\,\mathrm{fF}$, (b) short standard with $L_0=30\,\mathrm{pH}, L_1=10\times10^{-12}\,\mathrm{pH/Hz}, C_0=0.5\,\mathrm{fF}$, and (c) open standard with $C_0=15\,\mathrm{fF}, C_1=5\times10^{-15}\,\mathrm{fF/Hz}, L_0=1\,\mathrm{pH}$. All standards include a $200\,\mu\mathrm{m}$ CPW line offset. These models are used for automatic parasitic extraction.}
    \label{fig:4.2}
\end{figure}

The one-port SRM standards were implemented as non-ideal match, short, and open standards with a $200\,\mu\mathrm{m}$ offset, as shown in Fig.~\ref{fig:4.2}. For the network-load standards, a $4\,\mathrm{mm}$ line is used as the reciprocal standard, combined with the non-ideal match, short, and open standards. Note that the CPW has a complex-valued, frequency-dependent characteristic impedance that deviates slightly from $50\,\Omega$. Thus, the CPW line segments introduce both an offset and impedance mismatch, as $Z_\mathrm{ref}=50\,\Omega$ is used as the reference impedance for the lumped elements. The device under test (DUT) used for verification is a stepped-impedance line using the same CPW structure but with a $15\,\mu\mathrm{m}$ signal width.

Only the match and short standards have been used for automatic model parameter estimation. All lumped elements and material properties ($\epsilon_r^\prime$, $\tan\delta$, and $\sigma_r$) were assumed unknown. While including the open standard would have been possible, it would have increased the parameter space from 11 parameters (five from the match standard and six from the short standard) to 17 parameters, significantly increasing optimization time. Symmetric models at both ports were assumed to improve optimization efficiency, as asymmetric models would double the parameter count. This consideration is particularly important as the CPW model, though analytical, relies on computationally intensive Bessel functions. In the measurement example in Section~\ref{sec:4b}, devices have been modeled using only lumped elements without offset, allowing for more parameters and independent port solutions due to faster evaluation.

Though the CPW material parameters are common to both standards, they have been treated as independent during optimization to demonstrate the procedure's accuracy in parameter recovery.

To validate the accuracy of estimated parameters from the optimizer, a relative error metric has been defined:
\begin{equation}
	\text{Relative Error}(x)  = \left| \frac{x_\mathrm{est} - x_\mathrm{true}}{x_\mathrm{true}} \right|
	\label{eq:4.1}
\end{equation}
where $x_\mathrm{est}$ represents the estimated quantity and $x_\mathrm{true}$ is the true value. The estimated quantity corresponds to the parameter value retrieved from the optimization, while the true value is the actual value used to generate the simulation data. The relative error provides a normalized error metric that accommodates the different units of the different parameters.

Data processing was implemented in Python using the \textit{scikit-rf} \cite{Arsenovic2022} and \textit{scipy} \cite{Virtanen2020} packages. The optimization was performed using the DE algorithm \cite{Storn1997,Virtanen2020}, chosen for its robustness in handling high-dimensional parameter spaces without requiring gradient information or precise initial guesses. Furthermore, the SciPy implementation of DE supports parallelization, significantly accelerating convergence on multi-core processors compared to other global optimization routines available in the library. 

\begin{figure}[th!]
	\centering
	\includegraphics[width=0.98\linewidth]{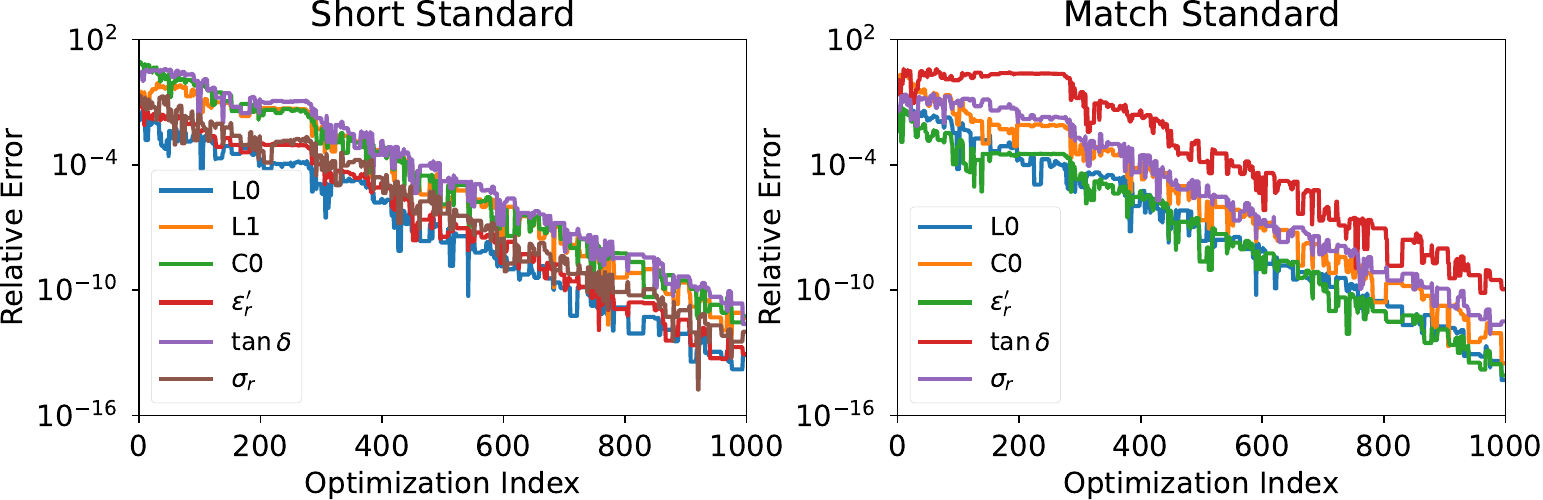}
	\caption{Relative errors in estimated parameters for short and match standards during DE optimization versus iteration count.}
	\label{fig:4.4}
\end{figure}

Fig.~\ref{fig:4.4} demonstrates the relative error of estimated parameters of the models used in SRM calibration, converging to floating-point precision. It should be noted that due to the computationally expensive CPW model, the 1000 iterations shown in Fig.~\ref{fig:4.4} took approximately 25 minutes to complete (using 150 frequency points).

\begin{figure}[th!]
    \centering
    \includegraphics[width=0.98\linewidth]{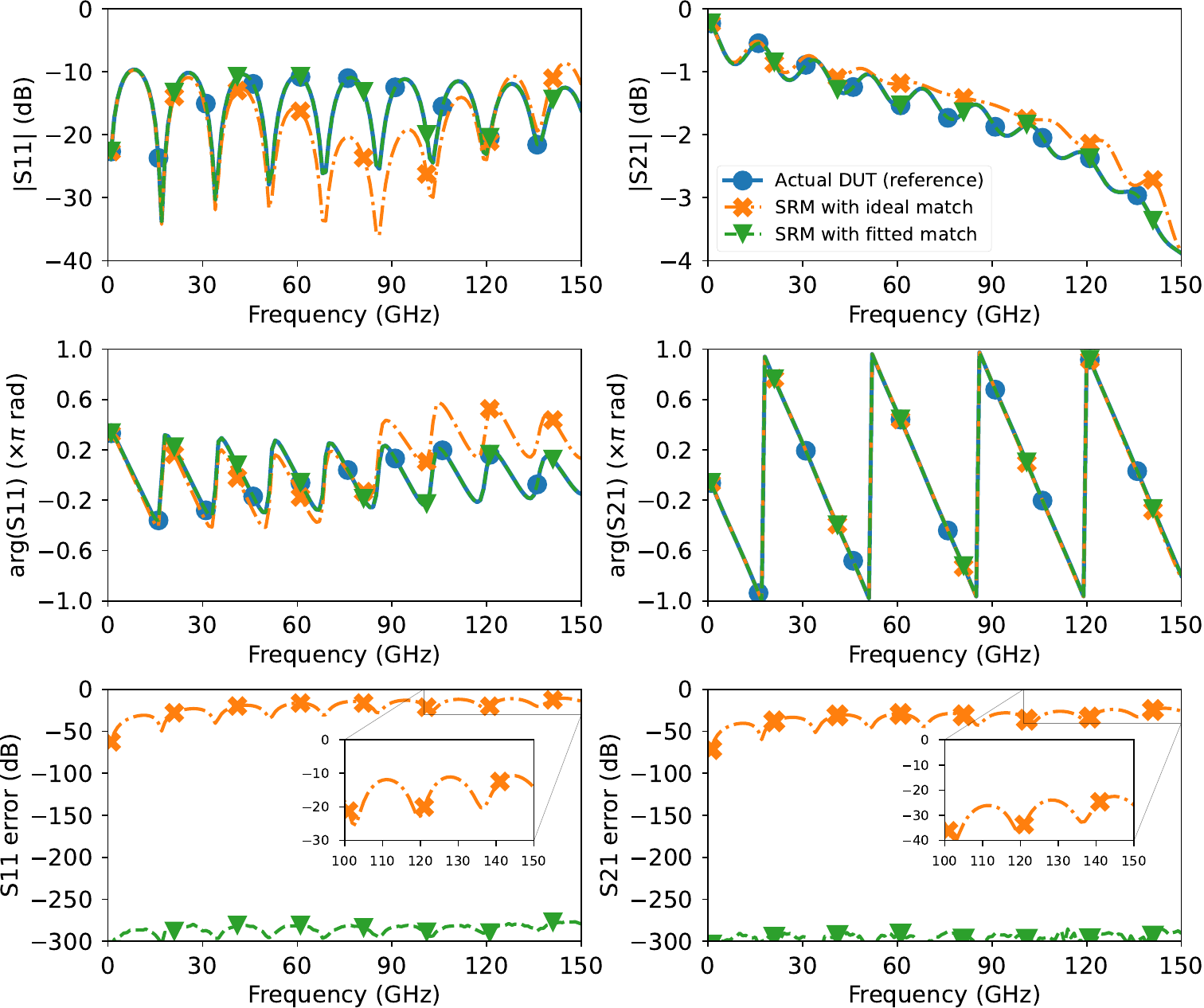}
    \caption{Comparison of the true S-parameters for the stepped impedance line (the actual simulated DUT prior to embedding with error boxes) against SRM calibration results obtained using the ideal match definition versus the model fitting procedure.}
    \label{fig:4.3}
\end{figure}

To compare S-parameters, absolute error is used to avoid division by zero, particularly for reflection parameters. The error is defined in dB as follows:
	\begin{equation}
		\text{Error}_{ij} (\mathrm {dB}) = 20\,\mathrm {log}_{10}\left |{ S_{ij}^{\mathrm {SRM}} - S_{ij}^{\mathrm {Ref}} }\right |
		\label{eq:4.2}
	\end{equation}
where $S_{ij}^{\mathrm {SRM}}$ denotes the S-parameter calibrated using the SRM method, and $S_{ij}^{\mathrm {Ref}}$ is the reference S-parameter. In this example, the reference corresponds to the actual simulated DUT before the error boxes are embedded. Fig.~\ref{fig:4.3} shows the calibrated DUT results comparing two cases: using an ideal match standard versus using nonlinear optimization to fit unknown parameters. This ideal match models the match response by a frequency-independent $50\,\Omega$ resistor, even if the match for the simulation is calculated based on the model introduced in Fig.~\ref{fig:4.2a}. The assumption of frequency independence of the match is well applicable for frequencies below $30\,\mathrm{GHz}$. Above this frequency, the ideal match standard definition produces significant errors in both $S_{11}$ and $S_{21}$. The optimization approach achieves much better accuracy, limited only by numerical precision. 

\subsection{PCB Measurement}
\label{sec:4b}
The experimental measurement compares multiline TRL calibration against SRM calibration with and without automatic model extraction. The calibration kits are fabricated on a PCB with microstrip interface. The substrate is based on ISOLA Tachyon-100G \cite{tachyon100g} with height of $100\,\mu\mathrm{m}$. The microstrip has a strip width of $230\,\mu\mathrm{m}$ and a copper thickness of $18\,\mu\mathrm{m}$. Measurements were conducted on a probe station using ground-signal-ground (GSG) probes with $150\,\mu\mathrm{m}$ pitch. The grounded CPW pads to microstrip transition design follows principles from \cite{Hatab2022a}.

For the multiline TRL standards, line lengths of $\{0, 0.5, 4, 5.5, 6.5, 8.5\}\,\mathrm{mm}$ were employed, with the shortest line serving as the thru standard. The reflect standard was implemented as an open circuit, achieved by separating the probes in air. The multiline TRL algorithm used to process the measurements is based on the method described in \cite{Hatab2022,Hatab2023}, which is als. 

\begin{figure}[th!]
    \centering
    \includegraphics[width=0.98\linewidth]{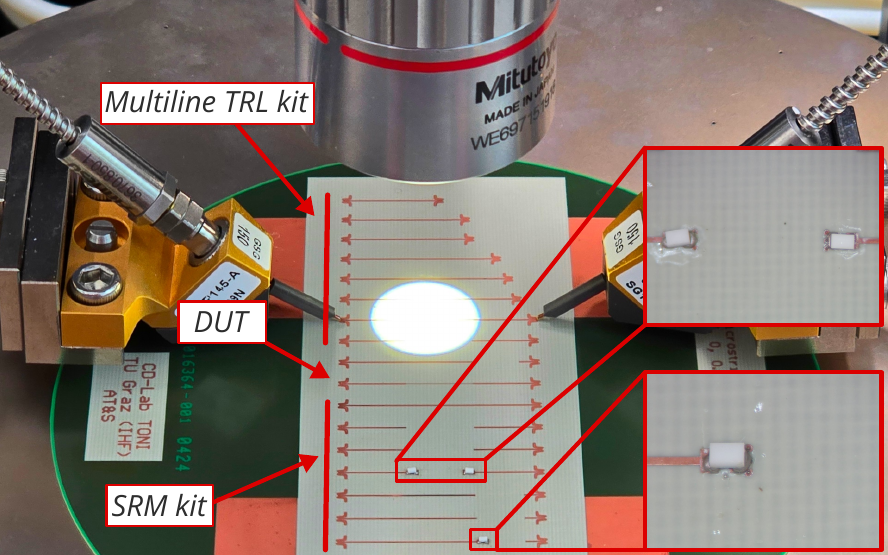}
    \caption{Probe station measurement setup showing the PCB containing both multiline TRL and SRM calibration kits and DUT. Inset photos show microscope views of the soldered flipchip resistors.}
    \label{fig:4.5}
\end{figure}

For SRM standards, symmetric open, short, and match standards, and network-load with an offset line segment were implemented. 
The match standard was implemented using a $50\,\Omega$ flip-chip Vishay CH0402-50RGFPA \cite{Vishay2020}, specified for operation up to $50\,\mathrm{GHz}$ with a $\pm2\%$ DC resistance tolerance. The DC resistance of the soldered flip-chip resistor was measured, indicating a value of $49\,\Omega$. This measurement was conducted by first zeroing the ohmmeter, which was connected through the bias-tee behind the probe at Port-A, and measuring a short standard implemented as a microvia. Subsequently, the soldered flip-chip resistor at Port-A was measured. The same measured DC value was used for port-B. The reference impedance of the SRM calibration was set to a constant reference impedance of $50\,\Omega$. It should be noted that due to the small size of the resistors (0402 footprint), manual soldering was challenging, potentially introducing asymmetry in frequency response due to the solder bumps. Fig.~\ref{fig:4.5} illustrates the measurement system and the soldered resistors.

\begin{figure}[th!]
    \centering
    \subfloat[]{\includegraphics[width=.5\linewidth]{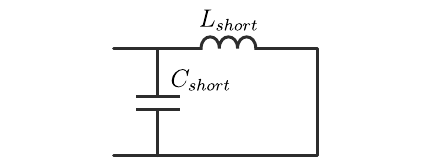}\label{fig:4.6a}}
    \subfloat[]{\includegraphics[width=.5\linewidth]{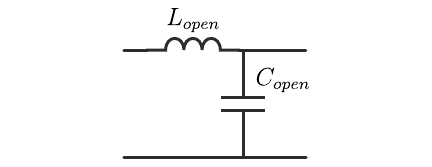}\label{fig:4.6b}}\\[-1pt]
    \subfloat[]{\includegraphics[width=.98\linewidth]{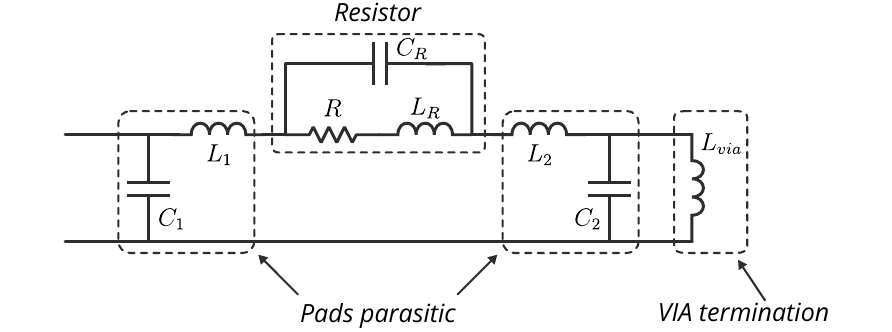}\label{fig:4.6c}}
    \caption{Equivalent circuit model for (a) short, (b) open, and (c) match standards on the PCB. No offset is considered as the calibration plane is located at the device.}
    \label{fig:4.6}
\end{figure}

For comparison between SRM and multiline TRL, a stepped impedance microstrip line was used as DUT with $100\,\mu\mathrm{m}$ strip width. For proper comparison, the multiline TRL calibration was renormalized to a constant reference impedance of $50\,\Omega$. Fig.~\ref{fig:4.6} shows the fitting models used for short, open, and match standards in SRM calibration. Both open and short standards were included in the optimization since they only comprise lumped elements and their evaluation is computationally efficient. The fitting was performed independently at each port. The total optimization time for both ports was approximately four minutes, fitting 22 parameters (using 197 frequency points). Only the parameters associated with the match results are used in the final SRM calibration. The resistor model follows the datasheet specifications \cite{Vishay2020}, with the microvia termination modeled as an inductor.

\begin{figure}[th!]
    \centering
    \includegraphics[width=0.98\linewidth]{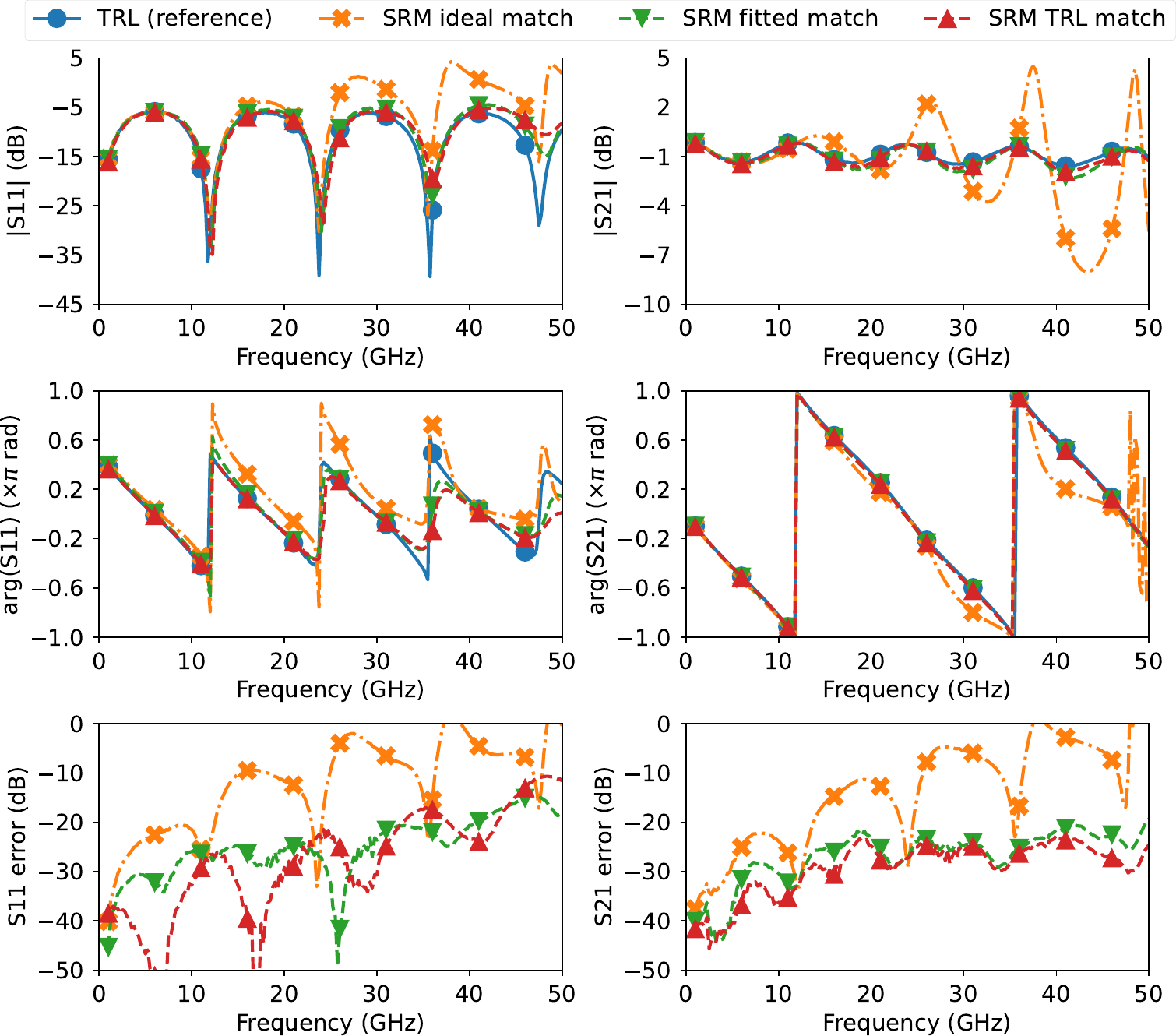}
    \caption{S-parameters comparison of the calibrated DUT, where the relative error is computed relative to the DUT calibrated with multiline TRL.}
    \label{fig:4.7}
\end{figure}

Fig.~\ref{fig:4.7} compares the S-parameters of the calibrated DUT between multiline TRL and SRM. The DUT response from multiline TRL calibration was used as the reference for calculating the absolute error based on \eqref{eq:4.2}. For SRM calibration, several scenarios were examined for defining the match standard: First, assuming ideal zero reflection (typical choice when frequency-dependent response is unknown); second, using our optimization procedure from Section~\ref{sec:3} to fit the models in Fig.~\ref{fig:4.6}; and finally, using the measurement of the match standard based on the multiline TRL calibration.

\begin{figure}[th!]
    \centering
    \includegraphics[width=0.98\linewidth]{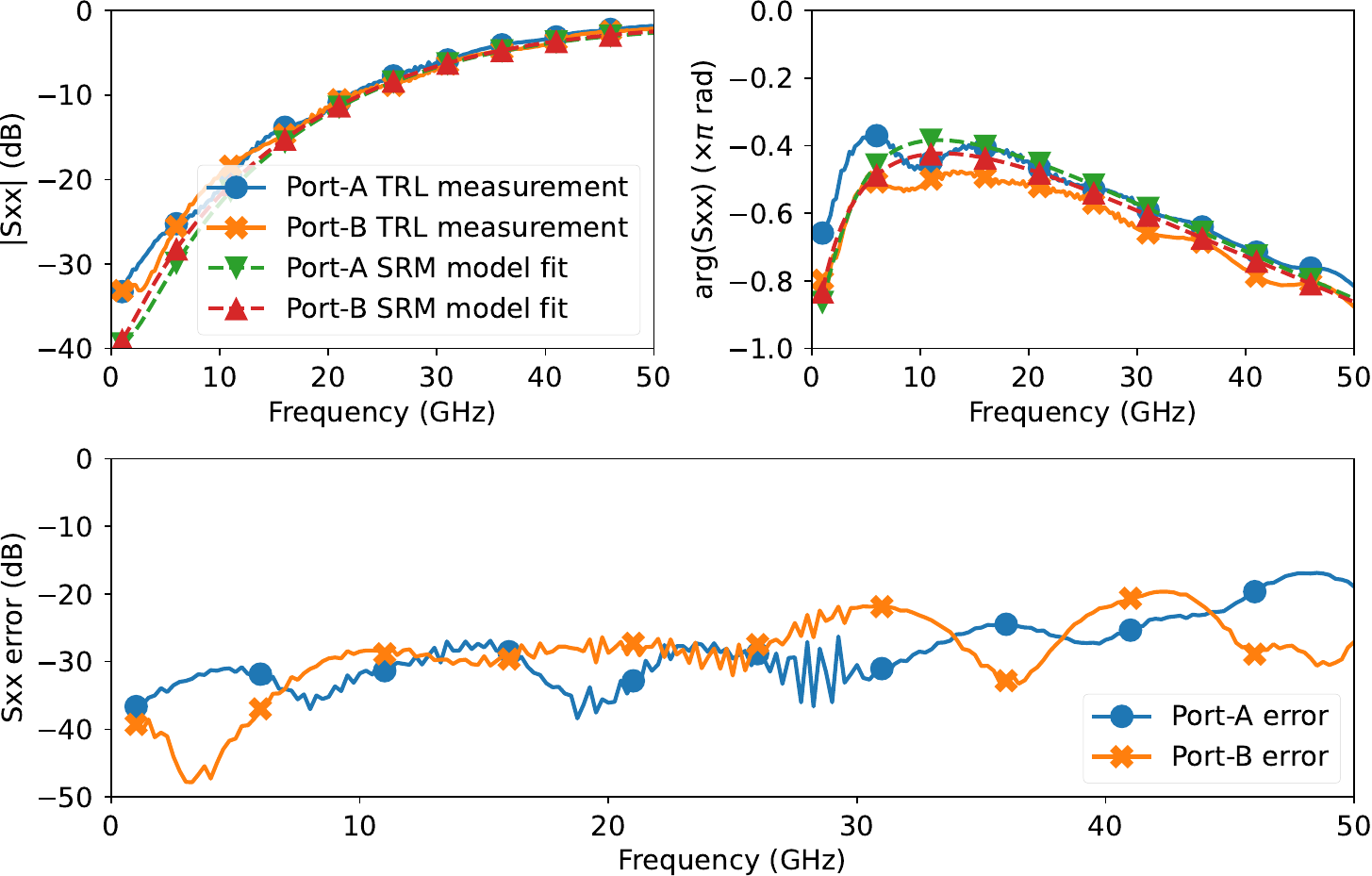}
    \caption{Calibrated reflection coefficient of the match standards comparing the SRM fit vs TRL measurement. Estimated parameters for the match standard model are listed in Table~\ref{tab:5.1}.}
    \label{fig:4.8}
\end{figure}

Results in Fig.~\ref{fig:4.7} demonstrate that assuming an ideal match response leads to substantial errors, particularly above $10\,\mathrm{GHz}$ in both $S_{11}$ and $S_{21}$. In contrast, the optimization approach yields errors comparable to defining the match standard based on multiline TRL. Some discrepancies exist between the DUT response based on SRM and multiline TRL calibrations, especially near $50\,\mathrm{GHz}$. This deviation may arise because the match standard exhibits behavior similar to a short standard near $50\,\mathrm{GHz}$, as evident in Fig.~\ref{fig:4.8}, thereby increasing the sensitivity of SRM calibration due to insufficient unique standards and causing the calibration problem to be ill-conditioned. Additionally, the discrepancy could stem from inconsistent resistor soldering, as asymmetry in the standards affects the calibration as investigated in \cite{Hatab2024}—an error that becomes amplified when the match standard behaves like a short standard near $50\,\mathrm{GHz}$. 

Nevertheless, achieving results comparable to multiline TRL, with an absolute error below $-22\,\mathrm{dB}$ for frequencies under $20\,\mathrm{GHz}$, is notable given the use of manually soldered flip-chip resistors on a PCB with significant parasitic responses. This is particularly significant considering only the DC resistance was specified during the SRM calibration. In contrast, assuming an ideal match yields poor results, exceeding $-10\,\mathrm{dB}$ at $20\,\mathrm{GHz}$ with significant degradation at higher frequencies. Furthermore, even for multiline TRL, the expected impedance variation of the traces can range from $\pm 4\,\Omega$ \cite{Hatab2023d}; hence, deviations between SRM and multiline TRL at lower frequencies are expected due to inherent impedance variations in PCB traces.

It is also worth noting that the SRM method with an optimized model yields a response similar to that obtained with SRM calibration and a defined match from multiline TRL measurements. This indicates that the deviation lies not in the definition of the match standard for the SRM method, but rather in the discrepancy between the standards. At higher frequencies, specifically above $20\,\mathrm{GHz}$, the match standard becomes increasingly reflective, as depicted in Fig.~\ref{fig:4.8}. Better agreement with multiline TRL at higher frequencies could be achieved by reducing resistor parasitics---for example, by using machine-placed resistors or higher-quality resistors where pad parasitics are less influential---or alternatively, by using multiple offset-short standards with different lengths to maintain a stable calibration problem at higher frequencies, as implemented in some coaxial calibration kits \cite{Blackham2003}.

Tables~\ref{tab:5.1} and \ref{tab:5.2} list the extracted model parameters for the match standard, open, and short standards from the SRM procedure for both ports. The similar values between port-A and port-B are due to the fact that these standards were designed to be symmetrical.
\begin{table}[ht!]
    \centering
    \caption{SRM estimated parameters for the match standard model based on schematic in Fig.~\ref{fig:4.6c}.}
    \label{tab:5.1}
    \begin{tabular}{@{$\,\,$}cccccccc@{$\,\,$}}
        \toprule
        & \begin{tabular}[c]{@{}c@{}}$C_1$\\ (fF)\end{tabular} & \begin{tabular}[c]{@{}c@{}}$L_1$\\ (pH)\end{tabular} & \begin{tabular}[c]{@{}c@{}}$C_2$\\ (fF)\end{tabular} & \begin{tabular}[c]{@{}c@{}}$L_2$\\ (pH)\end{tabular} & \begin{tabular}[c]{@{}c@{}}$C_R$\\ (fF)\end{tabular} & \begin{tabular}[c]{@{}c@{}}$L_R$\\ (pH)\end{tabular} & \multicolumn{1}{c}{\begin{tabular}[c]{@{}c@{}}$L_{via}$\\ (pH)\end{tabular}} \\ \midrule
        Port-A & 57.37 & 37.80 & 1.50 & 46.69 & 102.02 & 223.01 & 6.54 \\ \midrule
        Port-B & 53.68 & 50.60 & 1.59 & 28.95 & 108.55 & 216.46 & 4.52 \\ \bottomrule
    \end{tabular}
\end{table}

\begin{table}[ht!]
    \centering
    \caption{SRM estimated parameters for the short and open models based on schematic in Fig.~\ref{fig:4.6a} and~\ref{fig:4.6b}.}
    \label{tab:5.2}
    \begin{tabular}{@{$\qquad$}ccc|cc@{$\qquad$}}
        \toprule
        \multicolumn{1}{c}{} & \multicolumn{2}{c}{Short model} & \multicolumn{2}{c}{Open model} \\ \midrule
        &
        \begin{tabular}[c]{@{}c@{}}$C_{short}$\\ (fF)\end{tabular} &
        \begin{tabular}[c]{@{}c@{}}$L_{short}$\\ (pH)\end{tabular} &
        \begin{tabular}[c]{@{}c@{}}$C_{open}$\\ (fF)\end{tabular} &
        \begin{tabular}[c]{@{}c@{}}$L_{open}$\\ (pH)\end{tabular} \\ \midrule
        Port-A & 281.54 & 6.57 & 8.34 & 0.0095\\ \midrule
        Port-B & 199.67 & 6.48 & 8.73 & 0.0068\\ \bottomrule
    \end{tabular}
\end{table}


\section{Conclusion}
\label{sec:5}

In this paper, the capabilities of the SRM calibration procedure were extended to enable automatic frequency-dependent parasitic model extraction of the match standard during calibration. The model complexity can be arbitrarily defined by the user, provided that the DC value of the match is known and the number of unknown parameters is less than the number of frequency points. A global nonlinear optimization procedure finds the parameters that minimize the smallest singular value of the nullspace problem.

The capability of the SRM method was investigated to recover unknown parameters of the match standard through numerical simulation. The results demonstrated the ability to retrieve multiple parameters, including polynomial-based lumped elements and material properties of transmission line segments. Furthermore, the proposed approach was validated through measurements on a PCB using flip-chip resistors, which were modeled to account for both the non-ideality of the resistors themselves and the mounting pads, as well as the termination via arrays. The extracted model yielded results comparable to those obtained using a defined match standard measured with multiline TRL calibration.

The advantage of our proposed procedure over other calibration procedures that partially define the match standard, such as the LRRM method, lies in its modeling flexibility. While the LRRM method models the match using only an inductor, the SRM procedure can implement any combination of lumped elements or transmission lines, provided that the number of unknown parameters remains less than the frequency points. Furthermore, the SRM method offers additional advantages as all standards are defined through symmetry and reciprocity. With automatic model extraction of the match standard, only the DC resistance needs to be defined during calibration. However, one limitation of the SRM method compared to LRRM or LRM methods is the requirement for more partially defined standards, necessitating additional measurements.


\section*{Acknowledgment}
The authors thank AT\&S, Leoben, Austria, for the production of the PCBs used in this paper. 

\bibliographystyle{IEEEtran}
\bibliography{References/references.bib}

\begin{IEEEbiography}[{\includegraphics[width=1in,height=1.25in,clip,keepaspectratio]{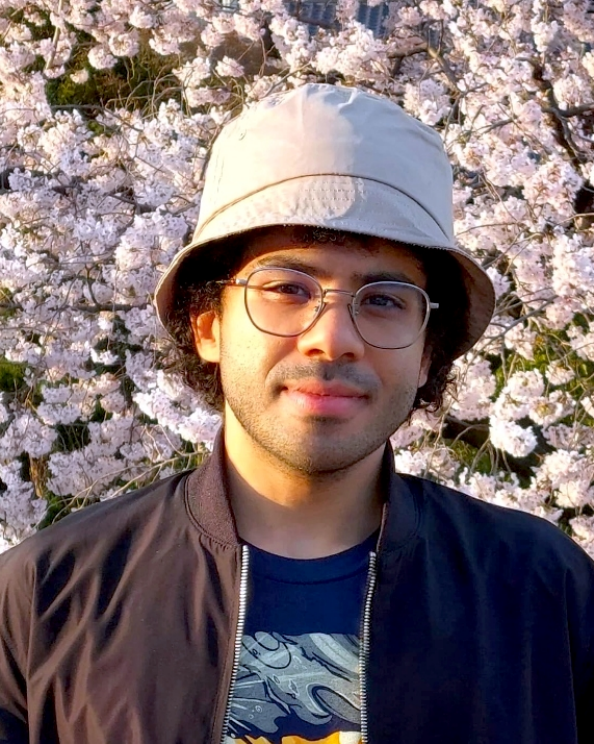}}]{Ziad Hatab} (Member, IEEE) received the Dipl.-Ing. and Dr.techn. degrees in electrical engineering from Graz University of Technology, Graz, Austria, in 2020 and 2024, respectively, at the Institute of Microwave and Photonic Engineering.

In 2020, he joined the Christian Doppler Laboratory for Technology-Guided Electronic Component Design and Characterization in Graz as a researcher. His research focused on vector network analyzer measurement techniques and calibration methods for planar circuit characterization at millimeter-wave and beyond. Since 2024, he has been an R\&D engineer at Keysight Technologies, Santa Rosa, CA, USA, where he works on vector network analyzer system design and advanced measurement techniques at millimeter-wave and sub-terahertz frequencies.
\end{IEEEbiography}

\begin{IEEEbiography}[{\includegraphics[width=1in,height=1.25in,clip,keepaspectratio]{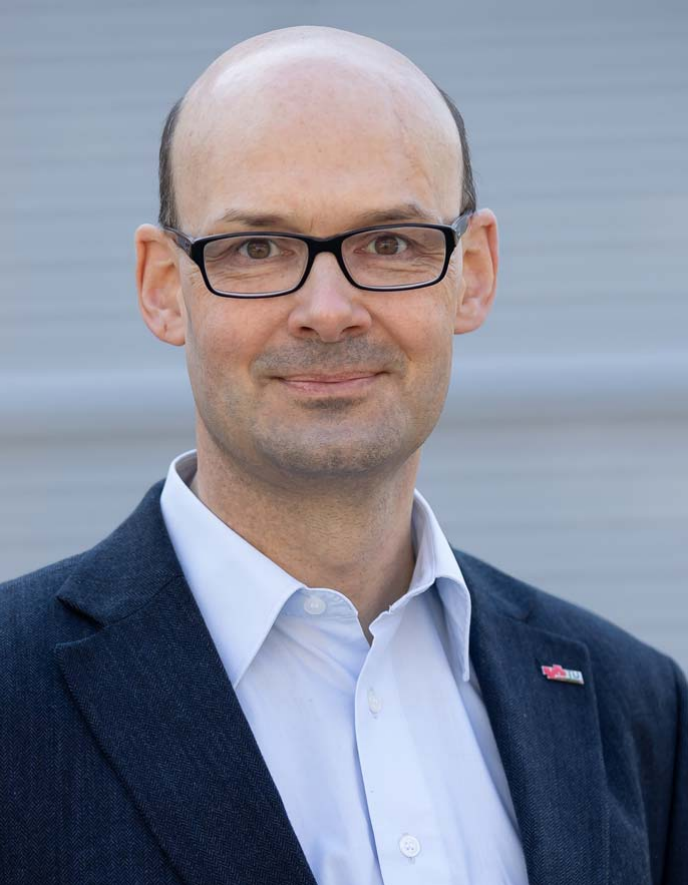}}]{Michael E. Gadringer}
Michael E. Gadringer (Senior Member, IEEE) is an Associate Professor at the Institute of Microwave and Photonic Engineering at Graz University of Technology, Austria. He received the Dipl.-Ing. and Dr. techn. degrees from Vienna University of Technology, Austria, in 2002 and 2012, respectively. In 2010, he changed to Graz University of Technology. From 2016 to 2021, Michael Gadringer held a tenure-track research and teaching position at this institute. He visited Rohde \& Schwarz GmbH in Munich in 2017 and Infineon Technology AG in 2018 in this context.

During his studies, he was involved in designing analog and digital linearization systems for power amplifiers and behavioral modeling of microwave circuits. His current research focuses on developing broadband microwave and mm-wave systems. Additionally, he is involved in planning and implementing complex measurements, with a focus on calibration techniques and material characterization. Michael Gadringer has authored more than 30 journal articles and 75 conference papers. He holds four worldwide patents and has co-edited the book “RF Power Amplifier Behavioral Modeling,” published by Cambridge University Press.

Michael Gadringer was a member of the IEEE 1765 standard working group on the recommended practice for estimating the Error Vector Magnitude of digitally modulated signals. In addition, he is contributing to the IEEE P2822 working group on the recommended practice for Microwave, Millimeter-wave, and THz on-wafer calibrations, deembedding, and measurements.
The IEEE Instrumentation and Measurement Society selected him as a 2020 IEEE TIM Outstanding Reviewer. Since August 2022, he has served as an Associate Editor of the IEEE Transactions on Instrumentation and Measurements.
\end{IEEEbiography}
\vfill

\begin{IEEEbiography}[{\includegraphics[width=1in,height=1.25in,clip,keepaspectratio]{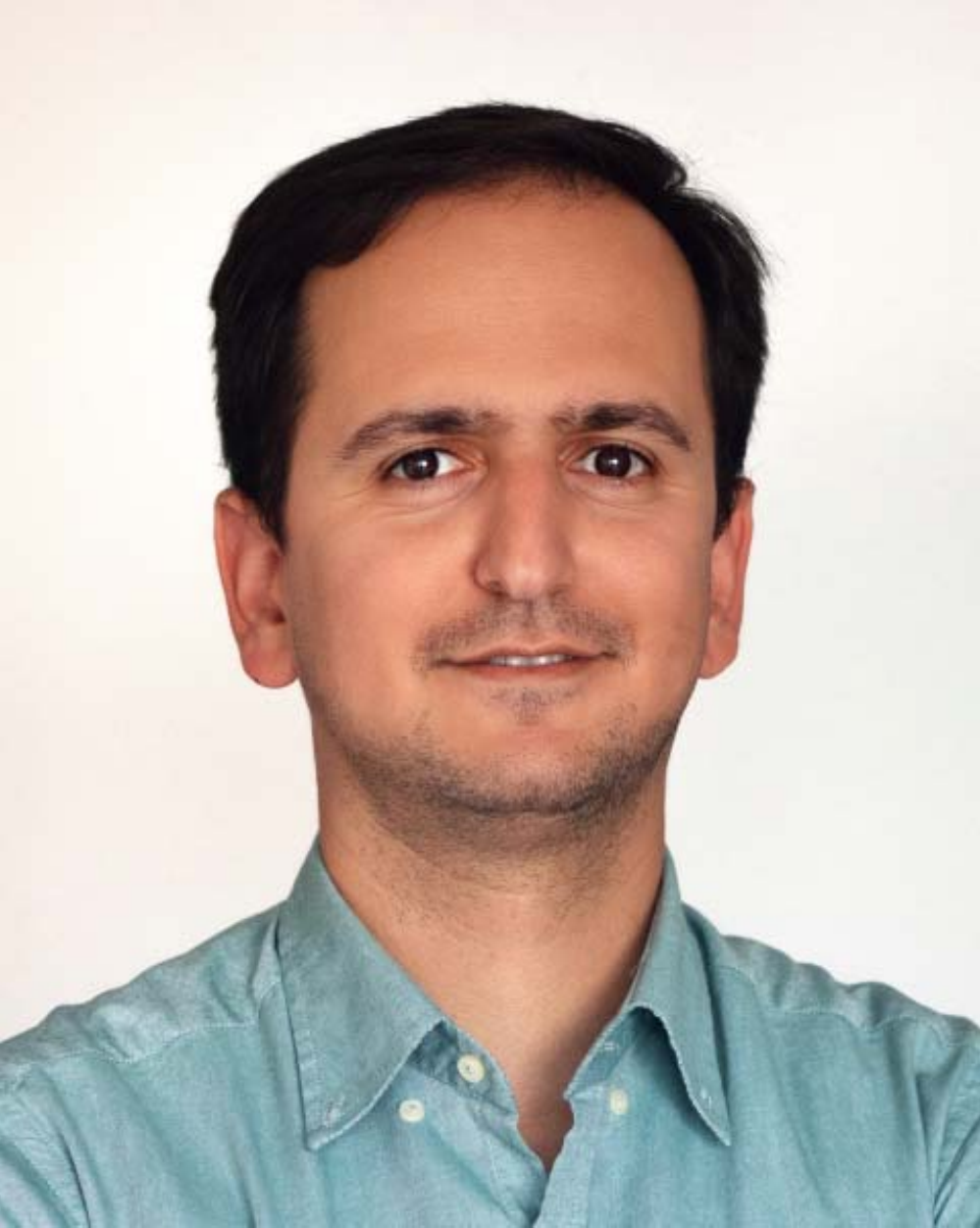}}]{Arash Arsanjani}
Arash Arsanjani (Member IEEE) was born in Kerman, Iran, in 1992. He received the Ph.D. degree in electrical engineering from the Graz University of Technology, Graz, Austria, in 2023. From 2019 to 2022, he was a Member of the European Union's Horizon 2020 research and innovation program for early-stage researchers, where his work focused on utilizing additive manufacturing for high-performance microwave and millimeter systems. Since 2022, he has been a University Project Assistant with the Christian Doppler Laboratory for Technology Guided Electronic Component Design and Characterization, Graz. His research interests include microwave and millimeter-wave passive components, filters, and antennas.
\end{IEEEbiography}

\begin{IEEEbiography}[{\includegraphics[width=1in,height=1.25in,clip,keepaspectratio]{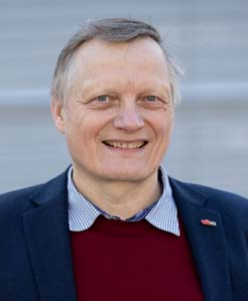}}]{Wolfgang Böosch}
Wolfgang B¨osch (Fellow, IEEE) received his Dipl.Ing. degree from the Technical University of Vienna, Austria, in 1985, his Ph.D. degree from the Graz University of Technology, Austria, in 1988, and his M.B.A. degree from the School of Management, University of Bradford, U.K., in 2004. In 2010, he joined Graz University of Technology to establish a new Institute for Microwave and Photonic Engineering. For nine years, he was the Dean of the Faculty of Electrical and Information Engineering, which currently incorporates 13 Institutes and 20 Full Professors covering the areas of Energy generation and distribution, Electronics, and Information Engineering. He is responsible for the strategic development, budget, and personnel of the Faculty.

Prior to this, he was the Chief Technology Officer (CTO) of the Advanced Digital Institute, Shipley, U.K. He was also the Director of Business and Technology Integration for RFMD, U.K. For almost ten years, he was with Filtronic plc, U.K., as the CTO of Filtronic Integrated Products and the Director of the Global Technology Group. Before joining Filtronic, he held positions with the European Space Agency (ESA), The Netherlands, working on amplifier linearization techniques, with MPR-Teltech, Burnaby, BC, Canada, working on MMIC technology projects, and with the Corporate Research and Development Group of M/A-COM, Boston, MA, USA, where he worked on advanced topologies for high-efficiency power amplifiers. For four years, he was with DaimlerChrysler Aerospace (now Hensoldt), Ulm, Germany, working on T/R modules for airborne radar.

He is a Fellow of the IEEE and the IET. He has published more than 180 papers and holds 4 patents.
\end{IEEEbiography}
\vfill

\end{document}